\newif{\ifjournal}
  \journalfalse

\ifjournal
  \documentclass{aa}
  \usepackage{graphicx,times,amssymb}
\else
  \documentclass{paper}
\fi

\renewcommand{\d}{\mathrm{d}}

\begin{document}

\title{Stacking Clusters in the ROSAT All-Sky Survey}
\ifjournal
  \author{Matthias Bartelmann and Simon D. M. White}
  \institute{Max-Planck-Institut f\"ur Astrophysik, P.O.~Box 1317,
    D--85741 Garching, Germany}
  \authorrunning{M. Bartelmann, S.D.M. White}
  \titlerunning{Stacking Clusters in the ROSAT All-Sky Survey}
\else
  \author{Matthias Bartelmann and Simon D. M. White\\
    Max-Planck-Institut f\"ur Astrophysik, P.O.~Box 1317, D--85741
    Garching, Germany}
\fi

\date{\em Astronomy \& Astrophysics, submitted}

\newcommand{\abstext}
 {Ongoing and planned wide-area surveys at optical and infrared
  wavelengths should detect a few times $10^5$ galaxy clusters,
  roughly $10\%$ of which are expected to be at redshifts
  $\gtrsim0.8$. We investigate what can be learned about the X-ray
  emission of these clusters from the ROSAT All-Sky Survey. While
  individual clusters at redshifts $\gtrsim0.5$ contribute at most a
  few photons to the survey, a significant measurement of the mean
  flux of cluster subsamples can be obtained by stacking cluster
  fields. We show that the mean X-ray luminosity of clusters with mass
  $M\gtrsim2\times10^{14}\,h^{-1}M_\odot$ selected from the Sloan
  Digital Sky Survey should be measurable out to redshift unity with
  signal-to-noise $\gtrsim10$, even if clusters are binned with
  $\Delta z=0.1$ and $\Delta\ln M\sim0.3$. For such bins, suitably
  chosen hardness ratio allows the mean temperature of clusters to be
  determined out to $z\sim0.7$ with a relative accuracy of $\Delta
  T/T\lesssim0.15$ for $M>10^{14}\,h^{-1}M_\odot$.}

\ifjournal
  \abstract{\abstext%
    }
\else
  \begin{abstract}
    \abstext
  \end{abstract}
\fi

\maketitle

\section{Introduction}

With moderately deep, wide-area imaging surveys in the optical or near
infrared it is now possible to detect large samples of galaxy
clusters. Dalcanton (1996) proposed that clusters could be detected as
surface brightness enhancements even when all but a few of their
galaxies are too faint to be detected individually.  Her suggested
procedure consists of identifying and removing stars and galaxies from
carefully flat-fielded images, smoothing the residual with a kernel
matched to the core size of clusters, and searching for significant
peaks in the resulting smoothed map.  Gonzalez et al.~(2001)
successfully constructed the Las Campanas Distant Cluster Survey
(LCDCS) by applying this technique to drift-scan data taken with the
Las Campanas Great Circle Camera (Zaritsky, Schectman \& Bredthauer
1996). They mapped well over 100 square degrees and constructed a
catalog of 1073 groups and clusters. The estimated redshift limits of
the catalog range from $\sim0.3$ for groups to $\sim0.9$ for massive
galaxy clusters.

The high intrinsic uniformity of drift-scan surveys like the LCDCS
makes them ideal for applying Dalcanton's cluster-detection
technique. In a theoretical study, Bartelmann \& White (2002) showed
that massive galaxy clusters should be detectable in the Sloan Digital
Sky Survey (SDSS) out to redshifts of $\sim1.2$ if data in the $r'$,
$i'$ and $z'$ bands are summed. For the final projected SDSS survey
area of $10^4$ square degrees, $\gtrsim10^5$ galaxy clusters should be
detectable at the 5-$\sigma$ level, and $\sim10\%$ of those are
expected to be at redshifts $\gtrsim0.8$.

Relatively little is known about the X-ray emission of clusters at
redshifts beyond $\sim0.5$ despite numerous cluster surveys based on
X-ray data. The main reason for this is the steep decrease with
redshift of the observed X-ray flux, which implies that at $z>0.5$
individual massive clusters produce at most a few photons in surveys
like the ROSAT All-Sky Survey (RASS; Snowden \& Schmitt 1990).
Detections of distant galaxy clusters from the X-ray data alone are
limited to very luminous systems which can be detected in the
restricted areas where deeper observations are available.

The upcoming availability of large cluster surveys in wavebands other
than the X-ray regime allows a reversal of the traditional X-ray
survey strategy. Rather than identifying clusters in the X-ray data,
it becomes possible to stack X-ray survey data for a large number of
fields where clusters are already known from other surveys. The low
background count rate at X-ray wavelengths makes this an efficient
technique for detecting the summed emission from a large stack of
clusters.

In this paper we investigate the prospects for using the RASS to
detect X-rays from suitable samples of clusters identified in the SDSS
data. In Sect.~\ref{sec:2} we describe our model for the cluster
population. This is based closely on the properties of nearby clusters
and specifies the cluster distribution in mass, redshift, X-ray
temperature and luminosity. In Sect.~\ref{sec:3} we convert cluster
X-ray luminosities to count distributions expected in the ROSAT
All-Sky Survey. Based on these, we calculate in Sect.~\ref{sec:4} the
expected signal-to-noise both for the detection of mean cluster
emission and for estimates of mean cluster temperature.
Sect.~\ref{sec:5} summarises and discusses our conclusions.

\section{Model specifications\label{sec:2}}

\subsection{Cosmology}

Much evidence suggests that the universe is spatially flat with low
nonrelativistic matter density $\Omega_0$. Baryons make up only a
small fraction of this matter; the rest is dark, presumably consisting
of some massive, weakly interacting particle. A cosmological constant
$\Omega_\Lambda$ or an equivalent ``quintessence'' field contributes
the remaining energy density. For definiteness, we assume
$\Omega_0=0.3$, $\Omega_\Lambda=0.7$ and $h=0.7$

We assume structure to form from an initially Gaussian density
fluctuation field $\delta$ with statistical properties specified by
its linear power spectrum, for which we adopt the CDM form given by
Bardeen et al.~(1986) with primordial spectral index $n=1$. The only
remaining free parameter is then the normalisation of the initial
fluctuation field which we take as $\sigma_8=0.9$. This value was
originally estimated based on the observed local abundance of galaxy
clusters (White, Efstathiou \& Frenk 1993; Eke, Cole \& Frenk 1996;
Viana \& Liddle 1996; Pierpaoli, Scott \& White 2001; Evrard et al
2002) but some recent analyses favour smaller values (Reiprich \&
B\"ohringer 2002; Viana, Nichol \& Liddle 2002; Lahav et al.~2002). We
will show results for $\sigma_8=0.9\pm0.1$.

\subsection{Cluster population}

Haloes form from Gaussian primordial density fluctuations through
gravitational collapse. Press \& Schechter (1974) first derived an
approximate formula for the mass distribution of haloes as a function
of redshift $z$. This has recently been modified by Sheth, Mo \&
Tormen (2001) and Sheth \& Tormen (2002) based on an ellipsoidal
rather than a spherical model for collapse. They give the differential
comoving number density of haloes as
\begin{equation}
  n(M,z)\,\d M=A\sqrt{\frac{2}{\pi}}
  \left(1+\frac{1}{\nu^{2q}}\right)\frac{\bar{\rho}}{M}
  \frac{\d\nu}{\d M}\exp\left(-\frac{\nu^2}{2}\right)\,\d M\;,
\label{eq:H1}
\end{equation}
where $\nu=\sqrt{a}\delta_\mathrm{c}\sigma^{-1}(M,z)$ defines the
linear amplitude required for collapse of a density fluctuation and
$\bar{\rho}$ is the mean cosmic density today. $\sigma(M,z)$ in this
definition is equal to $\sigma_0(M)D_+(z)$, where $\sigma_0(M)$ is the
present \emph{rms} fluctuation in the dark matter density contrast
within spheres containing the mean mass $M$, and $D_+(z)$ (with
$D_+(0)=1$) is the growth factor for the linear growing mode
(cf.~Carroll, Press \& Turner 1992). The linear density contrast
required for collapse $\delta_\mathrm{c}$ depends weakly on cosmology;
for the $\Lambda$CDM model we have chosen $\delta_\mathrm{c}=1.673$
(e.g.~{\L}okas \& Hoffman 2001). The parameters $A$, $a$ and $q$ are
constants; the original Press-Schechter formula is obtained from
(\ref{eq:H1}) by putting $A=0.5$, $a=1$ and $q=0$. This mass function,
with $A=0.322$, $a=0.707$ and $q=0.3$, has been shown to fit high
resolution numerical simulations of structure growth in a wide range
of cosmologies, provided the halo mass is defined at fixed density
contrast relative to the cosmic mean density (Jenkins et al.~2001).

Next, we need to know the X-ray luminosity of a cluster of mass
$M$. We adopt the observed relation between cluster temperature $T$
and bolometric X-ray luminosity $L_X$
\begin{equation}
  L_X=10^{44}\,\mathrm{erg\,s^{-1}}\,
  \left(\frac{kT}{1.66\,\mathrm{keV}}\right)^{2.331}\;,
\label{eq:H2}
\end{equation}
as derived by Allen \& Fabian (1998). Observations suggest that there
is little evolution in the $L_X-T$ relation out to redshifts
$z\sim0.4$ (e.g.~Mushotzky \& Scharf 1997; Allen \& Fabian 1998;
Reichart, Castander \& Nichol 1999). Lacking any reliable information
about evolution to higher redshifts, we assume (\ref{eq:H2}) to hold
at all redshifts.  This, of course, is a major uncertainty of our
study.

According to the virial theorem, a halo of mass $M$ in equilibrium at
redshift $z$ with a structure similar to observed clusters should have
a mean temperature given by
\begin{equation}
  kT=4.88\,\mathrm{keV}\,
  \left[\frac{M\,h(z)}{10^{15}M_\odot}\right]^{2/3}\;,
\label{eq:H3}
\end{equation}
where $h(z)$ is the Hubble constant at redshift $z$ in units of
$100\,\mathrm{km\,s^{-1}\,Mpc^{-1}}$ and, in contrast to
Eq.~(\ref{eq:H1}), $M$ is here defined as the mass interior to a
sphere with mean overdensity $200$ times the \emph{critical} value at
redshift $z$. Recall that we assume $h(0)=0.7$ throughout our
analysis. The constant in this relation is taken from the cluster
simulations of Mathiesen \& Evrard (2001; their Table 1) and is
appropriate for specifying the temperature of the best fit single
temperature model for the X-ray spectrum over the mass and redshift
ranges of interest. When necessary, we use an NFW model of
concentration parameter 5 to convert between cluster masses defined at
different overdensities.

\subsection{X-ray emission}

We assume that clusters emit X-rays through thermal radiation
according to a Raymond-Smith plasma model (Raymond \& Smith
1977). Apart from cluster temperature and redshift, the model has two
free parameters, the metal abundance and an overall normalisation
corresponding to the total X-ray luminosity. We fix the metal
abundance to $Z=0.3\,Z_\odot$ at all $z$ in agreement with the
observed abundances of local clusters (e.g.~Fukazawa et al.~1998). The
results of Schindler (1999) suggest little evolution towards higher
redshift and the final count rates we derive depend only very weakly
on metallicity. Thus neglecting any dependence on redshift does not
induce significant uncertainty.

Let $F_\nu(T,z)d\nu$ be the total X-ray luminosity emitted in the
spectral interval $[\nu, \nu+d\nu]$ by a cluster of temperature $T$ at
redshift $z$. If the cluster is observed in an energy band bounded by
$E_1$ and $E_2>E_1$, only a fraction $f$ of its bolometric flux is
included in the bandpass, where
\begin{equation}
  f=\int_{E_1(1+z)}^{E_2(1+z)}F_\nu(T,z)\,\d\nu\,
    \left[\int_{0}^{\infty}F_\nu(T,z)\,\d\nu\right]^{-1}\;.
\label{eq:X1}
\end{equation}
Thus the band-limited flux $S_X$ is related to the bolometric X-ray
luminosity through
\begin{equation}
  S_X=\frac{f\,L_X}{4\pi D_\mathrm{L}^2(z)}\;,
\label{eq:X4}
\end{equation}
where $D_\mathrm{L}(z)$ is the luminosity distance from the observer
to redshift $z$. Note that this flux must still be modified to account
for foreground absorption.

We use version 11.1 of the \emph{xspec} software package (Arnaud 1996)
to tabulate $f$ for an observing band between $0.5$ and $2.4\,$keV,
for cluster temperatures between $0.5$ and $12\,$keV, and for
redshifts between 0 and 2. Interpolating within this table and using
Eqs.~(\ref{eq:H2}), (\ref{eq:H3}) and (\ref{eq:X4}), we can convert
cluster masses to cluster temperatures, X-ray luminosities, and
finally to unabsorbed fluxes in the observed energy range.

The azimuthally averaged X-ray surface brightness profile
$\Sigma(\theta)$ of galaxy clusters is often modelled using the
so-called beta profile (Cavaliere \& Fusco-Femiano 1978),
\begin{equation}
  \Sigma(\theta)=\Sigma_0\,\left[1+
    \left(\frac{\theta}{\theta_\mathrm{c}}\right)^2
  \right]^{-(3\beta-1/2)}\;,
\label{eq:X5}
\end{equation}
where $\theta_\mathrm{c}$ is an angular core radius, and the amplitude
$\Sigma_0$ is chosen to produce the required X-ray flux $S_X$. Based
on observation, we choose $\beta=2/3$ (e.g.~Mohr, Mathiesen \& Evrard
1999). For the linear core radius $r_\mathrm{c}$, we adopt the
relation
\begin{equation}
  r_\mathrm{c}=125\,\mathrm{kpc}\,h^{-1}\,\left(
    \frac{L_X}{5\times10^{44}\,\mathrm{erg\,s^{-1}}}
  \right)^{0.2}\;,
\label{eq:X6}
\end{equation}
where $L_X$ is the X-ray luminosity between $0.5$ and $2.4\,$keV.
This relation is a fair representation of at least some clusters with
luminosities within $10^{43-45}\,\mathrm{erg\,s^{-1}}$ (Jones et
al.~1998). Following Vikhlinin et al.~(1998), we assume that
(\ref{eq:X6}) does not evolve with redshift. The angular core radius
is then $\theta_\mathrm{c}=r_\mathrm{c}\,D^{-1}(z)$, where $D(z)$ is
the angular-diameter distance. In fact, Eq.~(\ref{eq:X5}) is a poor
fit to the profiles of many clusters, particularly those with strong
apparent cooling flows. This is not, however, of any great consequence
for our modelling since the RASS does not, in any case, resolve the
inner regions of most clusters.

Having fixed $\beta$, $S_X$ and the angular core radius
$\theta_\mathrm{c}$, the beta profile is normalised by
\begin{equation}
  \Sigma_0=\frac{S_X}{2\pi\theta_\mathrm{c}^2}\;.
\label{eq:X7}
\end{equation}

\section{Halo detection\label{sec:3}}

\subsection{Point-spread function}

The point-spread function $f(\theta;E,\phi)$ of the ROSAT-PSPC had
three components, a Gaussian kernel, Lorentzian wings, and a component
which falls off exponentially with the angular separation $\theta$
from the centre of the image (Hasinger et al.~1995). The parameters
for these components generally depend not only on photon energy $E$,
but also on $\phi$, the off-axis angle of the source.

\begin{figure}[ht]
  \includegraphics[width=\hsize]{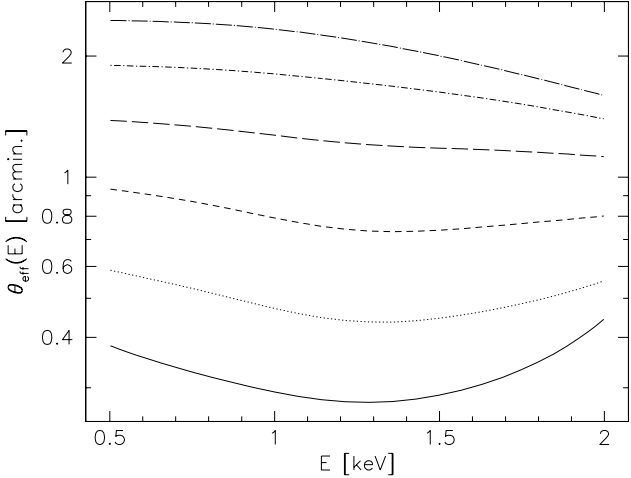}
\caption{Effective radii of the PSPC point-spread function as
  functions of photon energy, for off-axis angles between $10'$ and
  $60'$ (from bottom to top).}
\label{fig:4}
\end{figure}

The width of the PSPC point-spread function can be characterised by
the effective solid angle $\delta\Omega(E,\phi)$ covered,
\begin{equation}
  \delta\Omega(E,\phi)=2\pi\int_0^\infty\theta\d\theta\,
  f(\theta,E,\phi)\;,
\label{eq:R2}
\end{equation}
and we can define an effective radius $\theta_\mathrm{eff}(E,\phi)$ by
\begin{equation}
  \theta_\mathrm{eff}(E,\phi)=
  \left(\frac{\delta\Omega(E,\phi)}{\pi}\right)^{1/2}\;.
\label{eq:R3}
\end{equation}
The effective radii for six different off-axis angles between $10'$
and $60'$ are shown as functions of photon energy in Fig.~\ref{fig:4}.

\begin{figure}[ht]
  \includegraphics[width=\hsize]{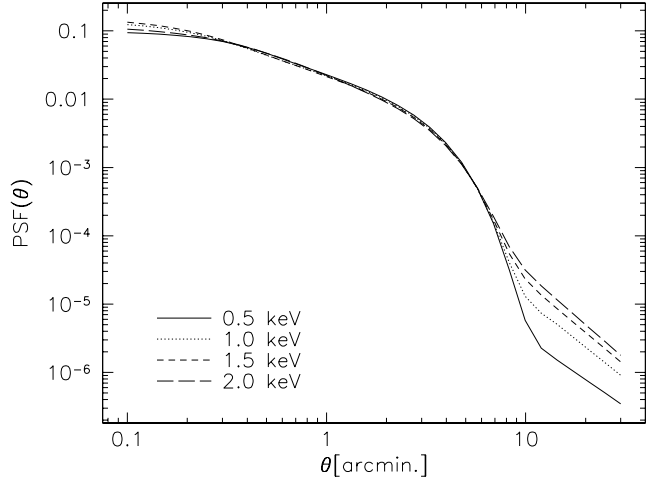}
\caption{The ROSAT-PSPC point-spread function, averaged over off-axis
  angles within the PSPC field-of-view. The different curves show the
  point-spread function for four different photon energies, as
  indicated.}
\label{fig:3}
\end{figure}

The field-of-view of the PSPC was large, with a radius of
approximately $60'$. Since a given point on the sky was scanned at
many different off-axis angles during the All-Sky Survey, the
appropriate point-spread function for the ROSAT All-Sky Survey at a
given photon energy is an area-weighted average of $f(\theta;E,\phi)$
over the field-of-view,
\begin{equation}
  \bar{f}(\theta,E)=\frac{2}{(60')^2}\,\int_0^{60'}\phi\d\phi\,
  f(\theta,E,\phi)\;.
\label{eq:R1}
\end{equation}
Figure~\ref{fig:3} shows the result for four different photon energies
between $0.5$ and $2.0\,$keV.

Figure~\ref{fig:3} shows that the point-spread function, averaged over
off-axis angles, falls to $\sim10\%$ of its peak value within 2--3 arc
minutes with little dependence on photon energy. Figure~\ref{fig:4}
confirms this weak dependence on photon energy, and illustrates the
strong dependence of effective PSF radius on off-axis angle. While
$\theta_\mathrm{eff}$ is below $0.5'$ for nearly on-axis photons, it
increases above $1'$ for photons coming from the edge of the
field-of-view.

The effective radius of the averaged point-spread function
$\bar{f}(\theta,E)$ can finally be averaged over photon energies to
obtain an average effective radius valid for the hard band of the
All-Sky Survey. Performing this average and weighting the photon
energies with the effective detector area as a function of $E$, we
find $\bar{\theta}_\mathrm{eff}=2.1'$.

\subsection{Converting fluxes to count rates}

We now need to estimate the signal expected in the All-Sky Survey from
a cluster with unabsorbed flux $S_X$ and temperature $T$ at redshift
$z$. To do this, we first modify the fluxes $S_X$ calculated using
Eq.~(\ref{eq:X4}) to allow for absorption by foreground neutral
hydrogen.  We assume a constant hydrogen column of
$4\times10^{20}\,\mathrm{cm}^{-2}$, which is typical for the high
galactic latitudes covered by the SDSS (e.g.~Dickey \& Lockman
1990). We convert the absorbed fluxes to PSPC count rates, using the
\emph{fakeit} task of the \emph{xspec} package with the PSPC response
matrix\footnote{electronically provided at
ftp://ftp.xray.mpe.mpg.de/rosat/calibration/data}.

In practice, we compute a two-dimensional table containing PSPC count
rates in the energy range between $0.5$ and $2\,$keV for
\emph{absorbed} Raymond-Smith spectra of a fixed \emph{unabsorbed}
flux normalisation and for cluster temperatures between $0.5$ and
$12\,$keV and redshifts between 0 and 2. Fluxes determined from
(\ref{eq:X4}) can then be converted to absorbed count rates by
interpolating within this table.

\subsection{Exposure times; background level}

\begin{figure}[ht]
  \includegraphics[width=\hsize]{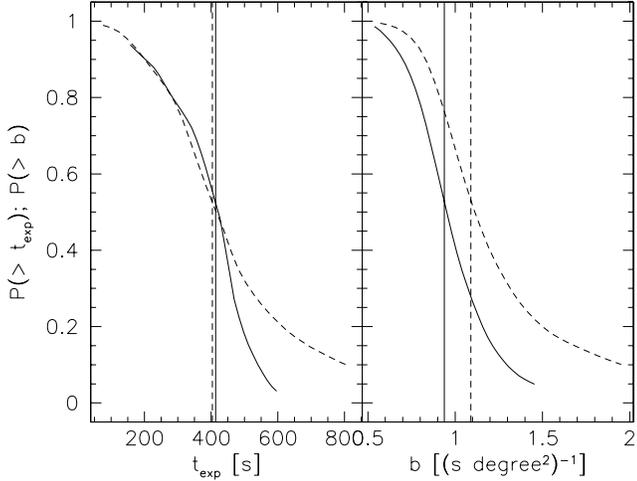}
\caption{Cumulative distributions of exposure-time (\emph{left panel})
  and background count-rate (\emph{right panel}) in the ROSAT All-Sky
  Survey. The dashed curves refer to the complete survey, the solid
  curves to the SDSS area only. The vertical lines mark the medians.}
\label{fig:1}
\end{figure}

The effective exposure time in the All-Sky Survey varies across the
sky because of the ROSAT scanning strategy. It is highest near the
ecliptic poles and lowest close to the ecliptic plane (cf.~Snowden et
al.~1995). Maps for the exposure time and the background count rates
were downloaded from the ROSAT All-Sky Survey web
page\footnote{http://www.xray.mpe.mpg.de/cgi-bin/rosat/rosat-survey/}.
The left panel in Fig.~\ref{fig:1} shows the cumulative exposure-time
distribution for the complete All-Sky Survey (dashed curve), and for
the area around the Northern Galactic cap covered by the Sloan Digital
Sky Survey. The median exposure times are marked by vertical lines.

The effective exposure times on the whole sphere and on the SDSS area
are only marginally different. For the SDSS area, we find a median
value
\begin{equation}
  t_\mathrm{exp}=414\,\mathrm{s}\;.
\label{eq:R4}
\end{equation}

Similarly, the background level is anisotropic across the sky. The
right panel in Fig.~\ref{fig:1} shows the cumulative distributions of
the background count rate in the All-Sky Survey for the whole sky
(dashed line) and for the SDSS area (solid line).

The background count rate within the SDSS area is noticeably lower
than on the whole sky; its median value is
\begin{equation}
  B=0.94\,\mathrm{s^{-1}\,deg^{-2}}=
  2.61\times10^{-4}\,
  \mathrm{s^{-1}\,arcmin^{-2}}\;.
\label{eq:R5}
\end{equation}

\section{Results\label{sec:4}}

\begin{figure}[ht]
  \includegraphics[width=\hsize]{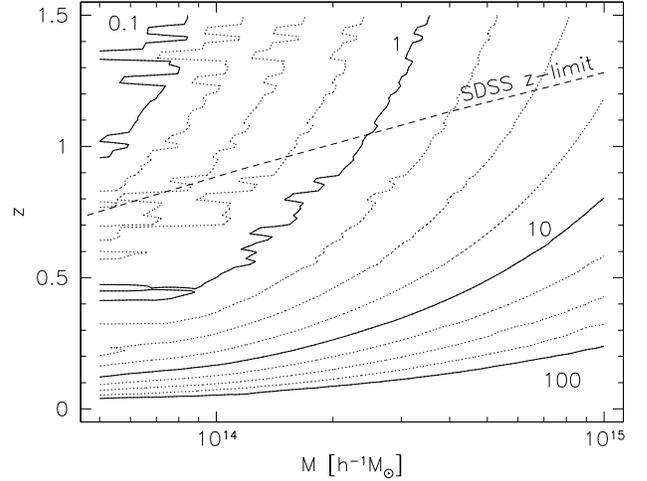}
\caption{Contours in the mass-redshift plane showing the counts
  received per cluster within the effective radius of the PSPC
  point-spread function. The contours are logarithmically spaced at
  0.25~dex between 0.1 (upper solid contour) and 100 counts (lower
  solid contour). The dashed curve marks the expected upper redshift
  limit for 5-$\sigma$ cluster detection in the combined $r'$, $i'$
  and $z'$ bands of the SDSS. The contours for low-mass clusters
  appear jagged because their X-ray spectra have strong features due
  to heavy elements which move relative to the observed energy band.}
\label{fig:6}
\end{figure}

Figure \ref{fig:6} shows photon-count contours in the plane spanned by
cluster mass and redshift. On a grid covering that plane, we compute
temperature, luminosity, flux, and count rate as described in the
previous section. We then multiply the count rate by the median
exposure time in the SDSS area, averaged over photon energies in the
$0.5-2.4\,$keV band.

The contours are logarithmically spaced by $0.25$ dex between 0.1 and
100 counts (upper and lower solid curves, respectively). They appear
jagged because a substantial fraction of the X-ray flux is contributed
by metal lines which move in and out of the observed energy band as
the redshift changes. The contours become smooth if the metal
abundance is set to zero. From this plot one can see, for example,
that the redshift limit below which individual clusters contribute
more than ten photons to the All-Sky Survey increases from
$z_\mathrm{max}\sim0.1$ at $M=10^{14}\,h^{-1}\,M_\odot$ to
$z_\mathrm{max}\sim0.8$ at $M=10^{15}\,h^{-1}\,M_\odot$. The dashed
curve shows the redshift limit for detection of clusters as 5-$\sigma$
surface brightness enhancements in the combined $r'$, $i'$ and $z'$
data of the SDSS (Bartelmann \& White 2002). Clearly the SDSS should
produce cluster catalogues which are much deeper at all masses than
those that can be made from the RASS.

Figure~\ref{fig:6} illustrates that only 0.3 photons per cluster are
expected for clusters of $M\sim10^{14}\,h^{-1}\,M_\odot$ at redshift
$z\sim0.8$. The number of such clusters expected in the SDSS is so
large, however, that it should be possible to determine their mean
X-ray properties by stacking data for many fields. This is true even
if the mass-redshift plane is divided into relatively narrow bins. We
now investigate this in more detail.

The background level of the All-Sky Survey is quite low, of order
$1\,\mathrm{s^{-1}\,deg^{-2}}$ which translates to approximately $0.8$
total counts per resolution element within the median exposure time of
the survey. The background will nevertheless dominate the noise in a
stacked image of distant clusters. Let $B$ be the mean surface density
of background photons in a single image, and $C(M,z)$ be the expected
number of photons from a single cluster of mass $M$ at redshift
$z$. Let $p(\theta)$ be the expected surface density of these cluster
photons as a function of angular distance $\theta$ from cluster
centre. $p(\theta)$ is given by a convolution of the mean cluster
surface brightness profile [Eq.~(\ref{eq:X5})] with the
point-spread-function of the survey (Fig.~\ref{fig:3}) and we
normalise it so that $\int p(\theta)\,2\pi\theta\,\d\theta=1$. In
practice for distant clusters the \emph{p.s.f.} is much broader than
the cluster image so that $p(\theta)$ is proportional to the
\emph{p.s.f.} itself.

\begin{figure}[ht]
  \includegraphics[width=\hsize]{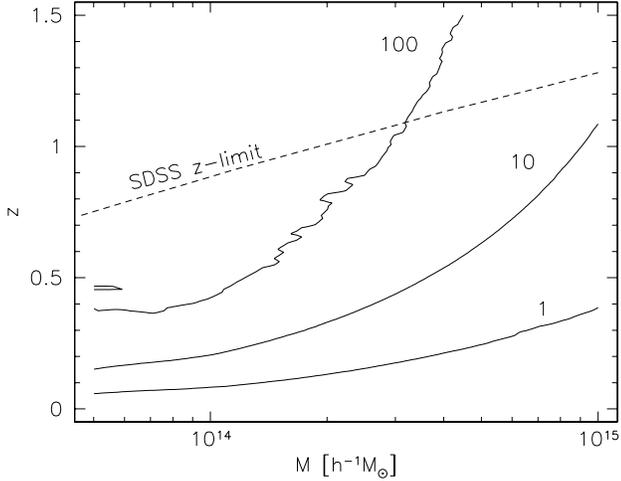}
\caption{Number of clusters to be stacked to achieve a 5-$\sigma$
  detection of their total X-ray emission in the ROSAT All-Sky
  Survey. The contour levels are 1 (lower contour), 10, and 100 (upper
  contour). The dashed curve marks the expected upper redshift limit
  for 5-$\sigma$ cluster detection in the combined $r'$, $i'$ and $z'$
  bands of the SDSS.}
\label{fig:7}
\end{figure}

For a stack of $N$ cluster fields the surface density of the
background is $NB$ and the expected surface density profile is
$NCp(\theta)$. Assuming Poisson photon statistics, the optimal
estimator of the cluster signal is then:
\begin{equation}
   \tilde{NC} = \int w(\theta)[O({\theta}) - B]2\pi\theta\d\theta\;,
\label{eq:estimdef}
\end{equation}
where $2\pi O(\theta)\theta\d\theta$ is the observed photon count in
an annulus width $\d\theta$, and the filter function $w$, normalised so
that $\int w(\theta)p(\theta)\,2\pi\theta\d\theta=1$, is given by
\begin{equation}
  w(\theta)=\frac{p(\theta)}{p(\theta)+B/C}\,
  \left[\int\frac{p^2 2\pi\theta\d\theta}{p+B/C}\right]^{-1}
\label{eq:filterdef}
\end{equation}
Clearly the expectation value of the estimator of equation
(\ref{eq:estimdef}) is just $NC$ while its variance is
\begin{equation}
  \mathrm{Var}(\tilde{NC})=NC\int w^2(\theta)p(\theta)\,
  2\pi\theta\d\theta\;.
\label{eq:varest}
\end{equation}
Thus the expected signal-to-noise for detecting the stacked cluster is
\begin{equation}
  \left(\frac{S}{N}\right)=(NC)^{1/2}\,
  \left[
    \int\frac{p^2(\theta)\,2\pi\theta\d\theta}{p(\theta)+ B/C}
  \right]^{1/2}\;.
\label{eq:S/N}
\end{equation}
If clusters are individually well above background ($Cp(\theta)\gg B$
over most of the broadened image) this gives the obvious result,
$(S/N)\approx(NC)^{1/2}$ for the stack. When background dominates
($Cp(0)\ll B$) the corresponding result is $(S/N)\approx NC/[NB\int
p^2(\theta)\,2\pi\theta\d\theta]^{1/2}$. In both cases the
signal-to-noise of the detection grows as $N^{1/2}$ for the stacked
image. Figure~\ref{fig:7} shows the number of cluster fields required
for a 5-$\sigma$ detection in the stacked image as a function of
cluster mass and redshift.

Contours are shown for $N=1$ (lower solid curve), $N=10$, and $N=100$
(upper solid curve). The figure shows that it takes 100 stacked
cluster fields to obtain a 5-$\sigma$ detection of clusters with
$M\sim10^{14}\,h^{-1}\,M_\odot$ at redshift $z\sim0.4$, but the
contours rise steeply enough that with the same number of stacked
fields one reaches redshifts above unity for cluster masses
$M\gtrsim3\times10^{14}\,h^{-1}\,M_\odot$. As in Fig.~\ref{fig:6}, the
dashed line shows the upper redshift limit expected for 5-$\sigma$
cluster detection in the combined $r'$, $i'$ and $z'$ bands of the
SDSS.

We now have to compare the number of cluster fields needed to achieve
a high signal-to-noise ratio with All-Sky Survey data to the number of
clusters we can expect to be available. The idea is to select fields
around clusters which are known from other data, and we continue to
take the SDSS as an example. We therefore ask how many clusters can be
expected in the SDSS data.

To give specific examples, we select two redshift intervals of width
$\Delta z=0.1$ each, one over $0.6\le z\le0.7$ and the other over
$0.9\le z\le1.0$. Our previous work has obtained the expected redshift
limit $z_\mathrm{lim}(M)$ as a function of cluster mass for detection
in SDSS data (Bartelmann \& White 2002). For a 5-$\sigma$ detection in
the combined $r'$, $i'$ and $z'$ bands, it is indicated by a dashed
line in Figs.~\ref{fig:6} and \ref{fig:7}. For each redshift interval,
we thus know the completeness limit in cluster mass, i.e.~the lowest
cluster mass $M_\mathrm{lim}$ above which clusters in that interval
are expected to be detectable. For the lower and upper redshift
intervals defined above, we obtain
$M_\mathrm{lim}=3.9\times10^{13}\,h^{-1}M_\odot$ and
$M_\mathrm{lim}=2.0\times10^{14}\,h^{-1}M_\odot$ respectively. For
each interval we then define a series of mass bins between
$M_\mathrm{lim}$ and $10^{15}\,h^{-1}M_\odot$ such that $\Delta\ln
M\sim0.3$.

\begin{figure}[ht]
  \includegraphics[width=\hsize]{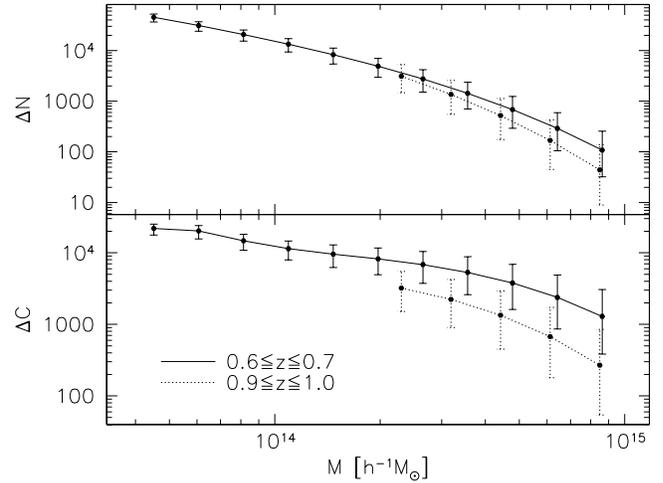}
\caption{Number of clusters $\Delta N$ (upper panel) and total cluster
  counts $\Delta C$ (lower panel) in the two redshift intervals
  $0.6\le z\le 0.7$ (solid curve) and $0.9\le z\le1.0$ (dotted curve)
  in mass bins of logarithmic width $\Delta\ln M\sim0.3$ between the
  SDSS completeness limit in the respective redshift interval and
  $10^{15}\,h^{-1}M_\odot$. The total counts received from all
  clusters per mass bin drop much less steeply than the cluster number
  because the number of counts received per cluster increases strongly
  with cluster mass. The error bars bracket results obtained by
  changing $\sigma_8$ by $\pm0.1$ and illustrate the sensitivity to
  the power-spectrum normalisation.}
\label{fig:9}
\end{figure}

The number of clusters in the redshift interval $[z_i,z_i+\Delta z]$
per mass bin $[M_j,M_{j+1}]$ is obtained through an integral of the
mass function (\ref{eq:H1}) multiplied by the comoving cosmic volume,
\begin{equation}
  \Delta N_{ij}=\int_{z_i}^{z_i+\Delta z}\d z\,
  \int_{M_j}^{M_{j+1}}\d M\,
  n(M,z)\,\left|\frac{\d V}{\d z}\right|\,(1+z)^3\;.
\label{eq:D5}
\end{equation}
The volume per unit redshift is
\begin{equation}
  \left|\frac{\d V}{\d z}\right|=\pi D^2(z)\,
  \left|\frac{\d D_\mathrm{prop}}{\d z}\right|\;,
\label{eq:D6}
\end{equation}
where $D$ is the angular diameter distance and $D_\mathrm{prop}$ the
proper distance. The factor $\pi$ instead of $4\pi$ accounts for the
fact that the SDSS only covers a quarter of the
sky. Figure~\ref{fig:9} shows the resulting cluster numbers $\Delta
N_{ij}$ and the total photon numbers $\Delta C_{ij}$ expected from
these clusters. The solid and dotted curves show results for the lower
and upper redshift bins, respectively. In order to illustrate the
sensitivity of the results to $\sigma_8$, the error bars mark the
range obtained for $\sigma_8=0.9\pm0.1$. The curves showing the total
photon counts received in each mass bin are flatter than those showing
the total cluster number because clusters with higher mass are more
X-ray luminous.

\begin{figure}[ht]
  \includegraphics[width=\hsize]{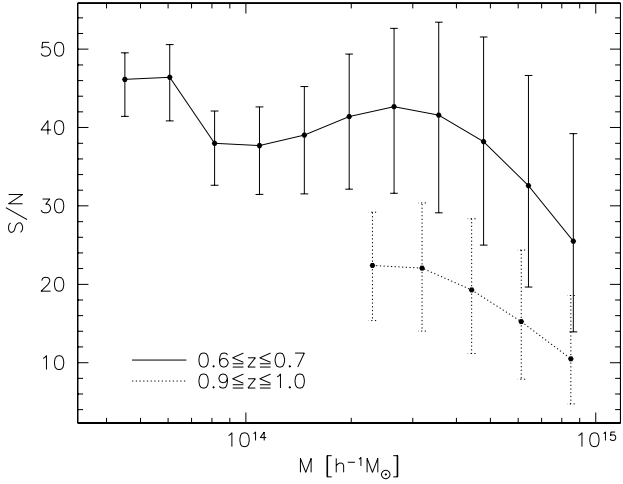}
\caption{Signal-to-noise ratios in stacked cluster fields in the given
  mass bins for the two redshift intervals $0.6\le z\le 0.7$ (solid
  curve) and $0.9\le z\le1.0$ (dotted curve). As in Fig.~\ref{fig:9},
  the error bars show the range obtained by varying $\sigma_8$ by
  $\pm0.1$. The signal-to-noise ratio in the lower redshift interval
  reaches $\sim40$ near $3\times10^{14}\,h^{-1}M_\odot$. Even near
  $10^{15}\,h^{-1}M_\odot$ in the upper redshift interval, the
  signal-to-noise ratio is $\sim10$.}
\label{fig:10}
\end{figure}

The figure shows that, even with relatively fine mass binning, more
than $10^4$ clusters should be detectable per mass bin below
$10^{14}\,h^{-1}M_\odot$ in the lower redshift interval $0.6\le
z\le0.7$. For comparison, Fig.~\ref{fig:7} shows that several hundred
stacked cluster fields are already sufficient for a 5-$\sigma$ X-ray
detection in the RASS. Similarly, more than $10^3$ clusters are
expected per mass bin below $4\times10^{14}\,h^{-1}M_\odot$ at higher
redshifts, $0.9\le z\le1.0$, where fewer than $\sim100$ cluster fields
need to be stacked for an X-ray detection. A useful way to quantify
these numbers is by calculating the expected signal-to-noise ratio in
a stack of all the cluster fields in each mass bin and in each of our
two redshift intervals. The results are shown in Fig.~\ref{fig:10}.

At the lower redshift the signal-to-noise ratio starts above $40$ near
$4\times10^{13}\,h^{-1}M_\odot$, where the contribution of metal lines
to the flux is high. With increasing mass, the line contribution
decreases and $S/N$ has a shallow minimum near
$10^{14}\,h^{-1}M_\odot$. Increasing continuum emission causes a broad
peak at $\gtrsim40$ centred on $3\times10^{14}\,h^{-1}M_\odot$. It
then decreases slowly towards higher masses. The drop-off results from
from the low cluster number at the high-mass end. If we set the metal
abundance to zero, the low X-ray flux at the low-mass end makes the
signal-to-noise drop to $\sim20$ near $4\times10^{13}\,h^{-1}M_\odot$.
Even in the upper redshift interval, the signal-to-noise ratio is
above 10, rising to $\gtrsim20$ in the lowest mass bin. These results
are, however, very sensitive to $\sigma_8$. Near
$10^{15}\,h^{-1}M_\odot$ in the upper redshift interval, the
signal-to-noise ratio varies between $\sim5$ and $\sim20$ as
$\sigma_8$ is increased from $0.8$ to $1.0$.

\begin{figure}[ht]
  \includegraphics[width=\hsize]{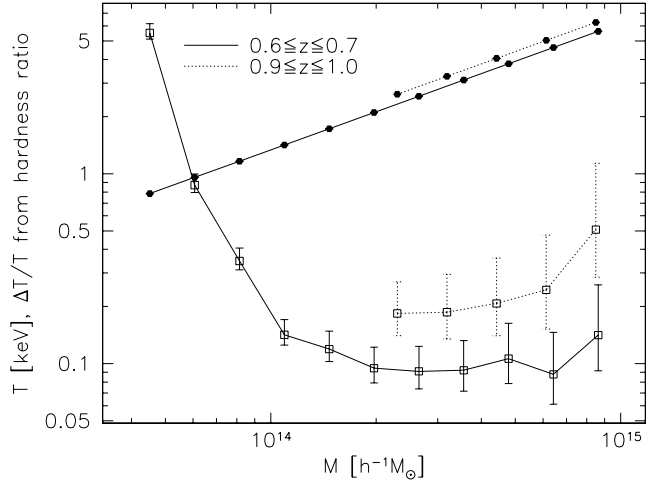}
\caption{The curves with open squares show the relative uncertainty
  $\Delta T/T$ of cluster temperatures determined from the hardness
  ratio between a soft ($E\in[0.5,1]\,$keV) and a hard
  ($E\in[1,2]\,$keV) band. Clusters are stacked in mass bins in the
  two redshift intervals $0.6\le z\le0.7$ (solid curve) and $0.9\le
  z\le1.0$ (dotted curve). As in Figs.~\ref{fig:9} and \ref{fig:10},
  the error bars indicate the range obtained by varying $\sigma_8$ by
  $\pm0.1$. While the temperature uncertainty is very large for the
  low-mass clusters, it drops near $10\%$ for moderate-redshift
  clusters with $M\gtrsim10^{14}\,h^{-1}M_\odot$, and is
  $\lesssim20\%$ for the high-redshift clusters with masses
  $\lesssim6\times10^{14}\,h^{-1}M_\odot$. The curves with filled
  circles show the cluster temperature in keV for the given mass bins
  and redshift intervals.}
\label{fig:11}
\end{figure}

The high signal-to-noise ratio even for high-redshift clusters
encourages us to investigate whether it will be possible to estimate
cluster temperatures from hardness ratios. We introduce two energy
bands, one with $0.5\le E/\mathrm{keV}<1$ and the second with $1\le
E/\mathrm{keV}\le2$. The counts $C_{1,2}$ in these two bands determine
the hardness ratio
\begin{equation}
  \mathcal{R}=\frac{\mbox{hard counts}}{\mbox{soft counts}}=
  \frac{C_2}{C_1}\;.
\label{eq:D7}
\end{equation}
We use \emph{xspec} to compute the hardness ratio $\mathcal{R}(T,z)$
expected for RASS data for clusters with temperature $T$ at redshift
$z$. For clusters of mass $M$ at redshift $z$, the uncertainty in the
temperature measurement is then
\begin{equation}
  \Delta T(M,z)=
  \left(\frac{\partial\mathcal{R}}{\partial T}[T(M),z]\right)^{-1}\,
  \Delta\mathcal{R}\;,
\label{eq:D8}
\end{equation}
where the uncertainty $\Delta\mathcal{R}$ of the measured hardness
ratio (\ref{eq:D7}) is determined by the count statistics.  The
boundaries of the energy bands were chosen so that $\mathcal{R}$ is
typically of order unity in the mass and redshift ranges considered
here. The signal-to-noise ratio of the hardness ratio
$\mathcal{R}/\Delta\mathcal{R}$ is $\gtrsim10$ for all cluster mass
bins in the redshift interval $0.6\le z\le0.7$, and is $\gtrsim8$ for
the bins in the redshift interval $0.9\le z\le1.0$. The derivative of
$\mathcal{R}$ with respect to $T$ is $\sim0.8$ for $T\sim1\,$keV and
falls to $\sim0.1$ the highest temperatures. As a result temperature
determinations should be most accurate for clusters with
$M\sim10^{14}\,h^{-1}M_\odot$; at lower masses, line emission in the
low-energy band dominates and the uncertainty $\Delta\mathcal{R}$
increases because of poor photon statistics in the high-energy
band. We show $T$ and $\Delta T/T$ in Fig.~\ref{fig:11} for the same
mass bins and redshift intervals used previously. For comparison, the
plot also gives the mean cluster temperature expected as a function of
mass in each redshift interval. Note that both $T$ and $\Delta T/T$
are emitted rather than observed values.

Figure~\ref{fig:11} shows that the relative uncertainty in the mean
temperature of the clusters in each mass bin is remarkably small for
$0.6\le z\le0.7$. For cluster masses $>10^{14}\,h^{-1}M_\odot$ it is
$\Delta T/T\lesssim0.15$. Over the mass range
$10^{14}-10^{15}\,h^{-1}M_\odot$, it appears that a $>10\,\sigma$
measurement of cluster temperature should be possible. As in
Figs.~\ref{fig:9} and \ref{fig:10}, the error bars in
Fig.~\ref{fig:11} indicate the range obtained by varying the
power-spectrum normalisation $\sigma_8$ by $\pm0.1$. For clusters in
the high-redshift band, $0.9\le z\le1.0$, the relative temperature
uncertainty increases both because of count statistics and because of
decreasing sensitivity of $\mathcal{R}$ to $T$. Despite this,
temperature measurements at $3$ to $10\,\sigma$ should be
possible. Note that a careful maximum likelihood measurement of $T$
would give results with somewhat higher significance than the simple
hardness ratio approach we have adopted here.

\section{Discussion\label{sec:5}}

Ongoing and planned wide-area surveys will detect tens of thousands of
galaxy clusters out to redshifts near and above unity. For example,
searching for surface-brightness enhancements in a smoothed stack of
the $r'$-, $i'$- and $z'$-band data of the Sloan Digital Sky Survey
should allow clusters of $5\times10^{13}\,h^{-1}M_\odot$ to be
detected out to $z\sim0.7$, while $z>1$ is reached for masses above
$\sim3\times10^{14}\,h^{-1}M_\odot$ (Bartelmann \& White 2002).

We have investigated here whether existing X-ray data can be used to
measure the X-ray emission of these clusters by stacking sufficiently
many fields. We assume clusters to be distributed in mass and redshift
as given by the numerical results of Jenkins et al.~(2001). Their
temperatures are taken to be proportional to $[M\,h(z)]^{2/3}$, with
the normalisation taken from the N-body/SPH simulations of Mathiesen
\& Evrard (2001). We adopt the observed low-redshift relation between
bolometric X-ray luminosity and temperature, and we assume that it
holds at all redshifts.  We model cluster X-ray surface-brightness
profiles by a beta profile, although this has little effect on our
results because most distant clusters are not resolved in the RASS. We
convert the bolometric X-ray luminosity into a count rate using the
\emph{xspec} software, assuming a Raymond-Smith plasma model with a
metallicity of $0.3$ solar and a foreground neutral-hydrogen column of
$4\times10^{20}\,\mathrm{cm}^{-2}$.

The only suitable survey of the X-ray sky is the ROSAT All-Sky Survey
(RASS). With its median exposure time of approximately $415$ seconds
and its effective detector area of $\sim230\,\mathrm{cm}^2$, it
detected $\sim10$ photons from a cluster of mass
$10^{14}\,h^{-1}M_\odot$ at $z\sim0.1$, and only about one photon from
a similar cluster at $z\sim0.5$. Since the effective angular
resolution of the RASS is $\sim2'$, cluster emission is typically
spread over an effective solid angle of $\sim14$~square
arcminutes. Due to the low background of the PSPC detector, only
$\sim1.5$ background photons are expected within this solid angle
during the median RASS exposure time. This corresponds to the number
of photons expected from a cluster with mass
$M\sim10^{14}\,h^{-1}M_\odot$ at redshift $z\sim0.35$, or with mass
$M\sim4\times10^{14}\,h^{-1}M_\odot$ at $z\sim1$. Thus stacked cluster
fields are background dominated at lower mass or higher redshift than
this.

Requiring a signal-to-noise exceeding 5, we find that 100 fields must
be stacked to get a significant detection of clusters with
$M\sim10^{14}\,h^{-1}M_\odot$ at $z\sim0.4$, or with
$M\sim3\times10^{14}\,h^{-1}M_\odot$ at $z\sim1.0$. A stack of ten
cluster fields should give a 5-$\sigma$ detection of massive clusters
with $M\sim10^{15}\,h^{-1}M_\odot$ at $z\sim1.1$, should any such
exist.

The number of clusters expected in wide-field surveys like the SDSS is
enormous and allows the detection of X-ray emission from even fairly
low-mass clusters out to surprisingly high redshift.  In the redshift
interval between $0.6$ and $0.7$, the surface-brightness technique of
Dalcanton (1996) should detect clusters in the SDSS data down to a
mass limit of $\sim3.9\times10^{13}\,h^{-1}M_\odot$.  If we bin the
clusters by mass into logarithmic bins with width $\Delta\ln M=0.3$,
the signal-to-noise ratio for the X-ray detection exceeds 35 near
$10^{14}\,h^{-1}M_\odot$, rises above 40 near
$3\times10^{14}\,h^{-1}M_\odot$ and drops to $\sim25$ at
$10^{15}\,h^{-1}M_\odot$. In the interval between redshifts $0.9$ and
$1.0$, the mass completeness limit for SDSS cluster detection
increases to $2\times10^{14}\,h^{-1}M_\odot$, but X-ray detections are
still possible with signal-to-noise ratios above 10 if clusters are
binned by mass as described.

We have also shown that the signal-to-noise ratio of the stacked
cluster images is high enough to divide the photons into two energy
bands, $E\in[0.5,1]$ and $E\in[1,2]$, and to estimate cluster
temperatures from the count ratio. In particular, for clusters with
$0.6\le z\le0.7$ and masses $>10^{14}\,h^{-1}M_\odot$, the hardness
ratio changes with cluster temperature sufficiently strongly for mean
cluster temperatures to be determined with a typical relative
uncertainty of $\Delta T/T\lesssim15\%$.

Of course, these results depend on the modelling assumptions we have
made. Their sensitivity to changes in the power-spectrum normalisation
$\sigma_8$ is shown in Figs.~\ref{fig:9}, \ref{fig:10} and
\ref{fig:11}, where the error bars bracket results obtained adopting
$\sigma_8=0.9\pm0.1$. Other critical assumptions are that the relation
between bolometric X-ray luminosity and temperature is independent of
redshift, and that the cluster temperature scales with cluster mass as
given by simulations. Our assumptions about cluster X-ray profiles are
less critical because of the low angular resolution of the RASS.

Using photometric redshifts for brightest cluster members, it should
be possible to determine redshifts for SDSS clusters with an accuracy
of $\Delta z\sim0.05$. On the other hand, at best very rough estimates
of cluster mass can be obtained from the optical data. The optically
selected clusters in a given redshift interval can be binned by
magnitude, and the study suggested here will then give relations
between optical luminosity and mean X-ray luminosity and
temperature. The latter can then be used to give an improved estimate
of mean cluster mass.

Wide-area surveys in the microwave regime will be carried out in the
near future which will detect of order one cluster per square degree
trough the thermal Sunyaev-Zel'dovich effect. The \emph{Planck}
satellite, for instance, due for launch in early 2007, is expected to
detect of order $30000$ galaxy clusters on the sky outside the
Galactic plane, approximately 10\% of which will be at redshifts
beyond $0.5$. Stacking these clusters in the same way as described
here, and combining their total integrated Compton-$y$ parameter with
their X-ray emission, will allow their total baryonic mass and perhaps
their temperatures to be constrained.

\end{document}